\documentstyle[multicol,aps,epsfig]{revtex}

\begin{document}
\draft
\title{Global versus local billiard level dynamics: The limits of universality}
\author{M. Barth, U. Kuhl, and H.-J. St\"ockmann}
\address{Fachbereich Physik, Universit\"at Marburg, Renthof 5, D-35032 Marburg, Germany}
\date{Phys. Rev. Lett. {\bf 82}, 2026 (1999)}
\maketitle

\begin{abstract}
Level dynamics measurements have been performed in a Sinai microwave billiard 
as a function of a single length, as well as in rectangular billiards with randomly 
distributed disks as a function of the position of one disk.
In the first case the field distribution is changed {\it globally}, 
and velocity distributions and autocorrelation functions are well 
described by universal functions derived by Simons and Altshuler.  
In the second case the field distribution is changed {\it locally}. 
Here another type of universal correlations is observed. 
It can be derived under the assumption that chaotic 
wave functions may be described by a random superposition of plane waves. 
\end{abstract}


\begin{multicols}{2}
In 1972 Edwards and Thouless noticed that the conductivity of a disordered 
system is closely related to the sensitivity of its eigenvalues on an external 
perturbation \cite{Edw72,Tho74}. For a ring with a perpendicularly applied 
magnetic field they conjectured that the conductivity $C$ is proportional 
to the averaged curvature of the eigenvalues, 
$C \sim \left< |\frac{\partial^2E_n}{\partial\varphi^2}|_{\varphi=0} \right>$, 
where $\varphi$ is the magnetic flux through the ring. 
In 1992 Akkermans and Montambaux showed 
that the conductivity may alternatively be expressed in terms of the 
eigenvalue velocities, 
$C \sim \left< |\frac{\partial E_n}{\partial\varphi}|^2 \right>$ \cite{Akk92}. 
This suggests to rescale the parameter and the eigenvalues by
\begin{equation} 
\label{xdef}
x = \frac{1}{\Delta} \left<\left|\frac{\partial E_n}{\partial\varphi}\right|^2\right>^{1/2} \: \varphi,
\qquad 
\epsilon_n(x) = \frac{E_n(\varphi)}{\Delta}, 
\end{equation}
where $\Delta$ is the mean level spacing. Szafer, Simons and Altshuler 
studied a number of parametric correlations of the rescaled 
eigenenergies \cite{Sza93,Sim93b}, in particular the velocity 
autocorrelation function 
\begin{equation}
c(x) = \left< 
\frac{\partial\epsilon_n(X+x)}{\partial X} 
\cdot
\frac{\partial\epsilon_n(X)}{\partial X} 
\right> 
- \left< \frac{\partial\epsilon_n(X)}{\partial X} \right>^2,
\end{equation}
originally introduced by Yang and Burgd\"orfer \cite{Yan92}, and conjectured 
a universal behavior as long as the so-called zero-mode approximation 
holds, i.e., in the range where the energy fluctuations show random matrix 
behavior. For the velocity distribution Simons and Altshuler found a 
Gaussian behavior \cite{Sim93b}. The same behavior 
has been obtained by a completely different approach starting from the 
analogy between the level dynamics of a chaotic system and the dynamics of 
a one-dimensional gas with repulsive interaction \cite{Yuk85,Kol94a}. 
In the region of onset of localization deviations from the Gaussian 
behavior are found \cite{Fyo95a}.

Since in the zero-mode approximation the energy correlations of a 
disordered system are identical to that of random matrices, it came as no 
surprise that the universal behavior of parametric correlations was found 
in billiard systems as well \cite{Sim93b}.
Universal behavior was observed also for the 
hydrogen atom in a strong magnetic field \cite{Sim93c}, conformally 
deformed \cite{Bru96} and ray-splitting billiards \cite{Hlu}, and in 
the acoustic spectra of vibrating quartz blocks \cite{Ber}.
In all cases the general features of the conjectured universal behavior 
had been reproduced reasonably well, but a number of significant 
discrepancies remained unexplained.

This was our motivation to study different types of billiard level dynamics 
a bit more detailed. All results to be presented below have been obtained 
in microwave billiards \cite{Ste95}. Here it is sufficient to note that for 
flat resonators the electromagnetic spectrum is completely equivalent to 
the quantum mechanical spectrum of the corresponding billiard, as long as 
one does not surpass the frequency $\nu_{max}=c/2h$, where $h$ is the 
resonator height. In the experiments we choose $h=$ 8~mm yielding a maximum 
frequency of 18.74~GHz. 

\begin{figure}[ht]
\begin{minipage}{8cm}
\psfig{file=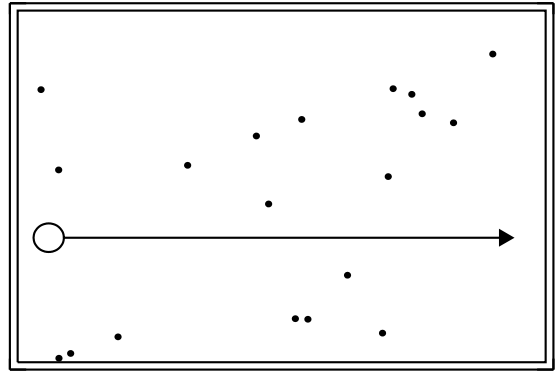,width=8cm}
\caption{\label{ScatBil}
Sketch of the billiard used for the local level dynamics (in scale).}
\end{minipage}
\end{figure}

One of the systems studied was a quarter Sinai billiard with a width 
$b=$ 200~mm, a radius $r=$ 70~mm of the quarter circle, and a length $a$ which 
was varied between 480 and 500~mm in steps of 0.2~mm. About 120 eigenvalues 
entered into the data analysis in the frequency range 14.5 to 15.5~GHz. 
The second system was a rectangular billiard with side lengths 
$a=$ 340~mm, $b=$ 240~mm, containing 20 randomly distributed circular disks 
with a diameter of 5~mm (see Fig.~\ref{ScatBil}). 
By a spatially resolved measurement \cite{ste92b} we found that all 
eigenfunctions $\psi$ in the studied frequency range were delocalized and  
$|\psi|^2$ was Porter-Thomas distributed. 
The position of one of the disks was varied in one 
direction in steps of 1~mm. Whereas the first type 
of level dynamics may be considered as global, since a shift of the 
billiard length of the order of 1 wavelength will change the wavefunction 
pattern everywhere in the billiard, the shift of the disk gives rise to a 
local modification only.

\begin{figure}[ht]
\begin{minipage}{8cm}
\psfig{file=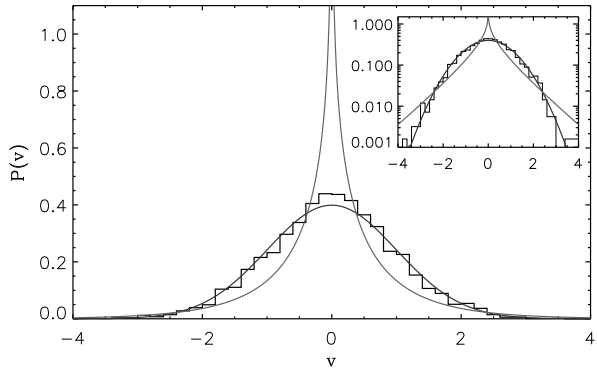,width=8cm}
\caption{\label{SinaiVel}
Velocity distribution in a quarter Sinai billiard with one length 
as the level dynamics parameter. 
The solid lines correspond to a Gaussian distribution and a distribution 
described by a modified Bessel function (see Eq.~(\ref{vlocal})), 
respectively. The inset shows the distribution in a logarithmic scale.}
\end{minipage}
\end{figure}

\begin{figure}[ht]
\begin{minipage}{8cm}
\psfig{file=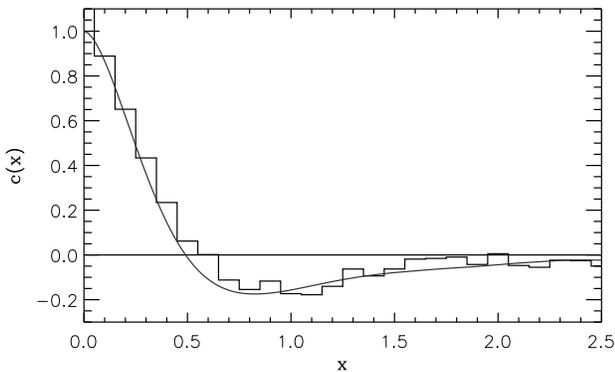,width=8cm}
\caption{\label{SinaiAC}
Velocity autocorrelation function in a quarter Sinai billiard, where 
the level dynamics parameter was scaled according to Eq.~(\ref{xdef}). 
No less than 2000 velocity pairs entered every bin of the histogram. 
The solid line corresponds to the universal autocorrelation function of 
Simons and Althsuler.}
\end{minipage}
\end{figure}

We start with a discussion of the global level dynamics. Figure \ref{SinaiVel} 
shows the velocity distribution for the quarter Sinai billiard with length 
$a$ as the level dynamics parameter. The distribution is well described by 
a Gaussian in accordance with the expected universal behavior (this result 
has been presented already in \cite{Kol94a}). Figure \ref{SinaiAC} shows the corresponding 
velocity correlator. To obtain the result, each eigenvalue was studied over 
a range of four to five avoided crossings, and the scaling was performed 
by calculating the mean squared velocity for each eigenvalue independently. 
Subsequently the results of about 120 eigenvalues were superimposed. 
The solid line corresponds to Simons' and Altshuler's universal function \cite{Muc}. 
The overall agreement between experiment and theory is good, but for $x > 2.5$ 
(not shown) the correlation function does not approach zero but stays at negative 
values. This is an artifact resulting from an 
insufficient number of data points making the calculation of the average 
$\left< \frac{\partial\epsilon_n(X+x)}{\partial X} \frac{\partial\epsilon_n(X)}{\partial X} \right>$ 
unreliable for large $x$ values. Most correlation functions found in the 
literature end at $x$ values of at most 1.5, probably just for this reason. 

\begin{figure}[ht]
\begin{minipage}{8cm}
\psfig{file=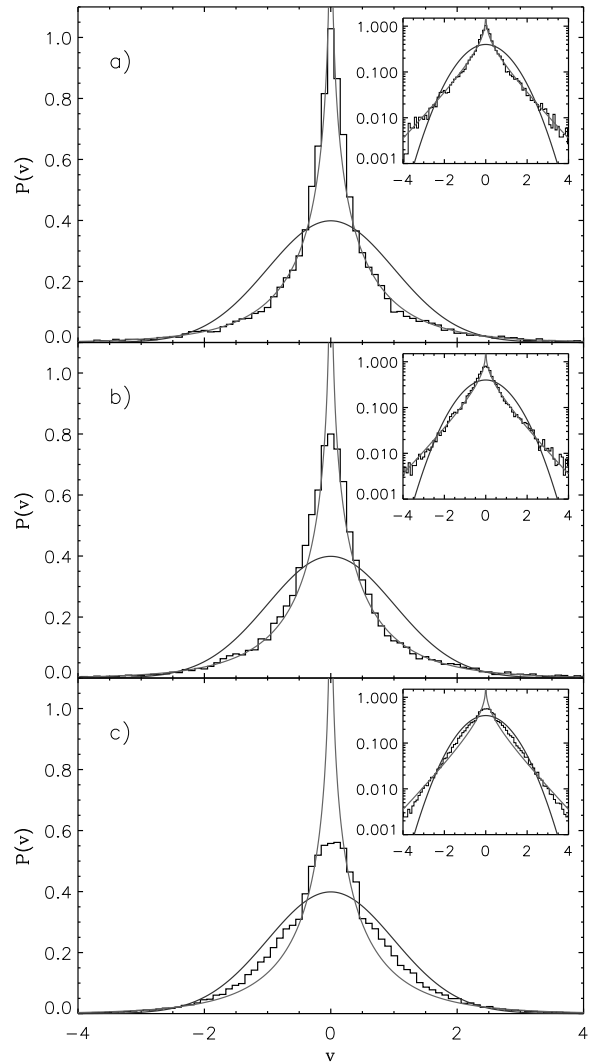,width=8cm}
\caption{\label{ScatVel}
Velocity distributions in a rectangular billiard with randomly 
distributed disks with the position of one disk with diameter $D$ as the 
level dynamics parameter. The ranges of $\delta=kD$ 
are $0.35<\delta<0.65$ (a), $1.4<\delta<2.6$ (b), $5.1<\delta<5.9$ (c).}
\end{minipage}
\end{figure}

\begin{figure}[ht]
\begin{minipage}{8cm}
\psfig{file=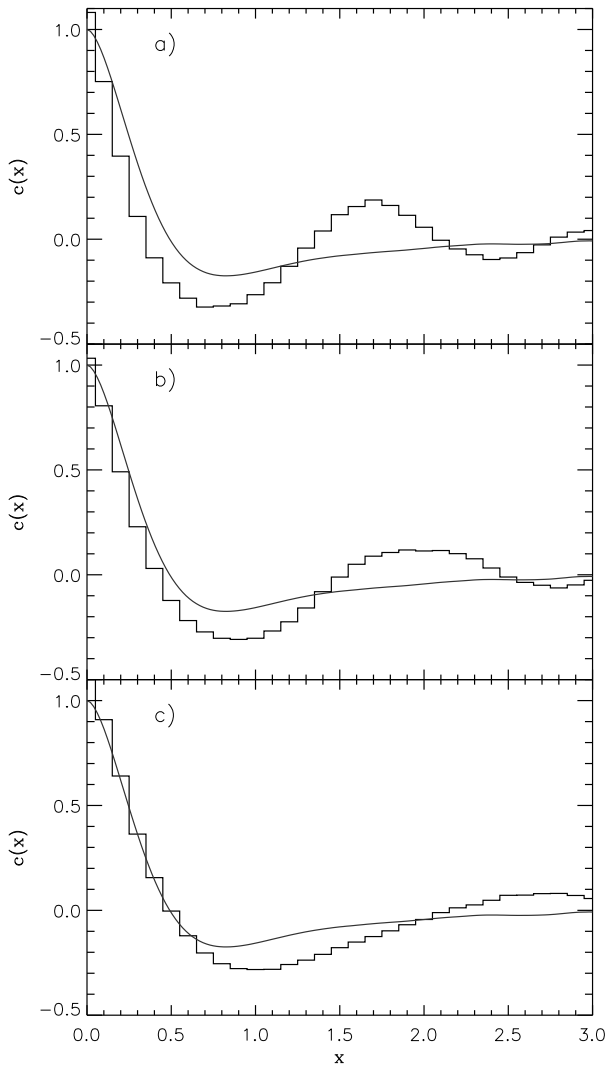,width=8cm}
\caption{\label{ScatACx}
Velocity autocorrelation functions in a rectangular billiard with 
randomly distributed disks, where the level dynamics parameter was scaled 
according to Eq.~(\ref{xdef}). The ranges of $\delta$ are the same as in 
Fig. \ref{ScatVel}.}
\end{minipage}
\end{figure}

Let us now turn to the discussion of the local level dynamics, where the 
position of one disk was varied. Whether a level dynamics must be 
considered as global or local, depends on the parameter $\delta=kD$, where 
$D$ is the diameter of the disk, and $k$ the wavenumber. It is well known 
that in the limit of small $\delta$ values the spectral properties of 
billiards containing hard spheres deviate significantly from random matrix 
behavior \cite{Dah98}.
Figure \ref{ScatVel} shows the velocity distributions for three different 
$\delta$ ranges. In Fig.~\ref{ScatVel}(a) a disk with $D=$ 5~mm was used, 
and the eigenvalues 
were taken in the frequency range 3.4 to 6~GHz. In Figs.~\ref{ScatVel}(b) 
and (c) the diameter of the movable disk was $D=$ 20~mm with eigenvalues 
in the frequency ranges 3.4 to 6 GHz and 12.5 to 14.5~GHz, respectively. 
None of the found velocity distributions is Gaussian. One observes instead a 
distribution with a pronounced peak at $v=0$, decreasing only exponentially 
for large values of $|v|$. With increasing $\delta$ values the distributions 
turn gradually into a Gaussian. We completed the series by a level dynamics 
measurement for a half Sinai billiard, where the position of the half 
circle was varied. Here the obtained velocity distribution (not shown),  
corresponding to $\delta$ values between 30 and 37, was already close to 
a Gaussian distribution. 

Figure \ref{ScatACx} shows the corresponding velocity 
autocorrelation functions. The scaling technique applied was the same as 
above. There is no longer any similarity between the experimental curves 
and the universal function. Only for the largest $\delta$ value displayed, 
the experimental curve seems to approach the Simons-Altshuler correlation 
function again. 

The results can be understood, if the movable disk is interpreted as a 
perturber probing the field in the resonator (the perturbing bead method 
has been used many years ago to map the field distributions in microwave 
cavities \cite{mai52a}, and has recently been applied to the study of 
wavefunctions in chaotic billiards as well \cite{sri91,doer98b,wu98}). 
In two-dimensional billiards the insertion of a metallic perturber leads 
to a negative frequency shift proportional to $E^2$, where $E$ is the 
electric field strength in the resonator in the absence of the perturber. 
This holds as long as the dimensions of the perturber are small 
compared to the wavelength, i.e., in the limit $\delta\to 0$. 
Applied to the present problem this means that the eigenvalue velocity is 
given by $\partial E_n / \partial r = \alpha\nabla|\psi|^2$ where $\nabla$ is the 
gradient in the direction of the displacement, and $\alpha$ is a constant depending on the 
geometry of the perturber. It follows for the velocity distribution function
\begin{equation}
\label{average}
P(v) = \left< \delta(v-2\alpha\psi\nabla\psi) \right>. 
\end{equation}

Under the assumption that the wavefunctions can be described 
by a random superposition of plane waves \cite{ber77a}, $\psi$ and 
$\nabla\psi$ are uncorrelated, and Gaussian distributed \cite{pri95b},
\begin{equation} 
\label{gaussian}
P_1(\psi) = \sqrt{\frac{A}{2\pi}} e^{-\frac{A\psi^2}{2}}, 
\qquad 
P_2(\nabla\psi) = \sqrt{\frac{A}{2\pi k^2}} e^{-\frac{A(\nabla\psi)^2}{2k^2}}. 
\end{equation}

The influence of the boundary is negligible here, since the linear dimensions 
of the billiard exceed the typical wavelength by factors of 5 to 10.
Using Eq.~(\ref{gaussian}) the average (\ref{average}) is easily calculated 
and yields
\begin{equation} 
\label{vlocal}
P(v) = \frac{\beta}{\pi} K_0(\beta|v|), 
\end{equation}
where $K_0(x)$ is a modified Bessel function, and $\beta=A/2\alpha k$. 
The solid lines plotted in addition to the Gaussian curves in 
Figs.~\ref{SinaiVel} and \ref{ScatVel} have been calculated from 
Eq.~(\ref{vlocal}). In the limit of small $\delta$ values distribution 
(\ref{vlocal}) describes the experimental distributions perfectly.

The influence of local perturbations on the energy levels has been 
studied by Aleiner and Matveev \cite{Ale98} who derived an explicit 
expression for the joint distribution function of initial and final energy 
levels. 
In their model the velocities are Porter-Thomas distributed \cite{Fyo}, if 
the coupling strength is taken as the level dynamics parameter.
The same distribution would have been expected in our case, if the coupling 
strength $\alpha$ would have been varied instead of the position (which, 
however, would be technically difficult to realize).

For the quadratical average of the eigenvalue velocities we obtain using 
Eq.~(\ref{vlocal})
\begin{equation} 
\left< \left( \frac{\partial E_n}{\partial r} \right)^2 \right> = \frac{1}{\beta^2}.
\end{equation}

Entering with this expression into Eq.~(\ref{xdef}), we get for the 
rescaled parameter
\begin{equation} 
\label{xlocal}
x = \frac{1}{\Delta\beta}r = \frac{\alpha}{2\pi}kr, 
\end{equation}
where we have used that in billiards the mean level spacing is given by 
$\Delta=4\pi/A$. Equation (\ref{xlocal}) shows that for the local level dynamics 
$x$ is not an universal parameter, since it depends via $\alpha$ on the 
geometry of the movable disk. We shall therefore use the rescaled parameter 
\begin{equation}
\label{barx}
\bar{x} = kr
\end{equation}
instead in the following. From the approach of random superposition of 
plane waves \cite{ber77a} the velocity autocorrelation function can be easily 
calculated, too. Using standard techniques as they are described, e.g., in 
Ref. \cite{Sre96b}, we get
\begin{equation} 
\label{clocal}
c(\bar{x}) = - \left[ J_0^2(\bar{x}) \right]'' 
= J_0^2(\bar{x}) - 2J_1^2(\bar{x}) - J_0(\bar{x}) J_2(\bar{x}). 
\end{equation}

\begin{figure}[ht]
\begin{minipage}{8cm}
\psfig{file=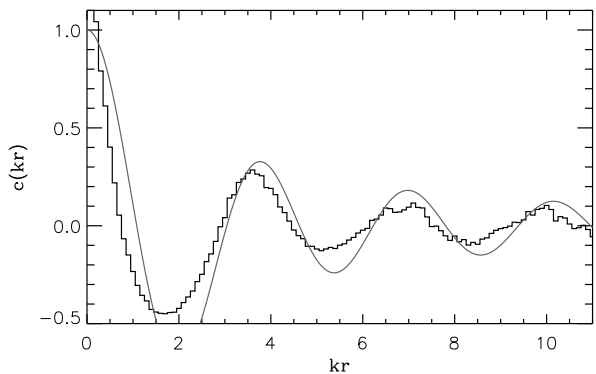,width=8cm}
\caption{\label{ScatACkr}
Same as Fig.~\ref{ScatACx}(a), but with the level dynamics parameter 
scaled according to Eq.~(\ref{barx}).}
\end{minipage}
\end{figure}

Figure \ref{ScatACkr} shows again the velocity autocorrelation of 
Fig.~\ref{ScatACx}(a) for the local level dynamics, but now as a function 
of $\bar{x}$. The solid line corresponds to the theoretial expectation (\ref{clocal}). 
The experimental curve follows closely the predicted oscillations. 
With increasing $\delta$ the oscillations are more and more damped, but the 
wavelength is still in accordance with the theory (not shown). 

This paper has shown that two different regimes of level dynamics have to be 
discriminated. In the local regime velocity distributions and 
autocorrelation functions are quantitatively described by the approach of
random superposition of plane wave, if the scaling (\ref{barx}) is applied. 
In the global regime, on the other hand, Simons' and Altshuler's universal 
functions describe the experimental results well, and the scaling (\ref{xdef}) 
is the appropriate one. The parameter $\delta=kD$ governs the transition 
between the two regimes.

Discussions with Y.V.~Fyodorov and T.~Guhr at different stages of the 
experiments are gratefully acknowledged. We thank H.U.~Baranger for calling 
our attention to Ref.~\cite{pri95b}, and E.R.~Mucciolo for making 
his calculations of the universal velocity correlator available to us.
The experiments were supported by the Deutsche Forschungsgemeinschaft 
via the SFB 185 "Nichtlineare Dynamik" and by an individual grant.

\end{multicols}


\begin{thebibliography}{10}
\bibitem{Edw72}
J.T. Edwards and D.J. Thouless, J. Phys. C {\bf 5}, 807 (1972).
\bibitem{Tho74}
D.J. Thouless, Phys. Rep. {\bf 13}, 93 (1974).
\bibitem{Akk92}
E. Akkermans and G. Montambaux, Phys. Rev. Lett. {\bf 68}, 642 (1992).
\bibitem{Sza93}
A. Szafer and B.L. Altshuler, Phys. Rev. Lett. {\bf 70}, 587 (1993).
\bibitem{Sim93b}
B.D. Simons and B.L. Altshuler, Phys. Rev. B {\bf 48}, 5422 (1993).
\bibitem{Yan92}
X. Yang and J. Burgd\"orfer, Phys. Rev. A {\bf 46},  2295  (1992).
\bibitem{Yuk85}
T. Yukawa, Phys. Rev. Lett. {\bf 54},  1883  (1985).
\bibitem{Kol94a}
M. Kollmann {\it et~al.}, Phys. Rev. E {\bf 49},  R1  (1994).
\bibitem{Fyo95a}
Y.V. Fyodorov and A.D. Mirlin, Phys. Rev. B {\bf 51}, 13403 (1995).
\bibitem{Sim93c}
B.D. Simons {\it et~al.}, Phys. Rev. Lett. {\bf 71}, 2899 (1993).
\bibitem{Bru96}
H. Bruus, C.H. Lewenkopf, and E.R. Mucciolo, Phys. Rev. B {\bf 53}, 9968 (1996).
\bibitem{Hlu}
Y. Hlushchuk {\it et~al.}, to be published.
\bibitem{Ber}
P. Bertelsen {\it et~al.}, to be published.
\bibitem{Ste95}
J. Stein, H.-J. St\"ockmann, and U. Stoffregen, Phys. Rev. Lett. {\bf 75}, 53 (1995).
\bibitem{ste92b}
J. Stein and H.-J. St\"ockmann, Phys. Rev. Lett. {\bf 68}, 2867 (1992).
\bibitem{Muc}
E.R. Mucciolo, private communication.
\bibitem{Dah98}
P. Dahlqvist and G. Vattay, J.~Phys.~A {\bf 31}, 6333 (1998).
\bibitem{mai52a}
L.C. Maier and J.C. Slater, J. Appl. Phys. {\bf 23}, 78 (1952).
\bibitem{sri91}
S. Sridhar, Phys. Rev. Lett. {\bf 67}, 785 (1991).
\bibitem{doer98b}
U. D\"orr {\it et~al.}, Phys. Rev. Lett. {\bf 80},  1030 (1998).
\bibitem{wu98}
D.H. Wu {\it et~al.}, Phys. Rev. Lett. {\bf 81}, 2890 (1998).
\bibitem{ber77a}
M.V. Berry, J.~Phys.~A {\bf 10}, 2083 (1977).
\bibitem{pri95b}
V.N. Prigodin {\it et~al.}, Phys. Rev. Lett. {\bf 75}, 2392 (1995).
\bibitem{Ale98}
I.L. Aleiner and K.A. Matveev, Phys. Rev. Lett. {\bf 80}, 814 (1998).
\bibitem{Fyo}
Y.V. Fyodorov, private communication.
\bibitem{Sre96b}
M. Srednicki and F. Stiernelof, J.~Phys.~A {\bf 29}, 5817 (1996).
\end{thebibliography}
\end{document}